\documentclass[12pt]{article}

\usepackage{amsmath,amssymb,amsthm,bm,braket,graphicx,mathpazo,mathrsfs}

\usepackage[final]{hyperref}

\usepackage[margin=1in]{geometry}

\usepackage{microtype}

\usepackage{mathtools}
\usepackage[english]{babel}

\usepackage[acronym]{glossaries}
\glsdisablehyper

\newacronym{ata}{ATA}{all-to-all}
\newacronym{kt}{KT}{kicked~top}
\newacronym{rmt}{RMT}{random matrix theory}
\newacronym{nns}{NNS}{nearest-neighbour spacing}
\newacronym{goe}{GOE}{Gaussian orthogonal ensemble}
\newacronym{gue}{GUE}{Gaussian unitary ensemble}
\newacronym{eth}{ETH}{eigenstate thermalisation hypothesis}

\newcommand{\myi}{\ensuremath{\mathrm{i} }}
\newcommand{\normaldist}{\mathscr{N}}

\numberwithin{equation}{section}

\usepackage{thmtools}

\theoremstyle{definition}

\newcommand{\sref}[1]{Section~\ref{#1}}

\newcommand{\tauata}{\tau_\text{A}}
\newcommand{\taukt}{\tau_\text{T}}
\newcommand{\alphaata}{\alpha}

\newcommand{\subata}{\text{A}}
\newcommand{\subkick}{\text{K}}
\newcommand{\subtop}{\text{T}}
\newcommand{\subchain}{\text{C}}
\newcommand{\hamZ}{\ensuremath{H_\text{K'}} }
\newcommand{\hamO}{\ensuremath{H_\text{T}} }

\newcommand{\spinsize}{J}
\newcommand{\spinnumber}{N}
\newcommand{\isingweight}{\ensuremath{\tau_\text{C} }}
\newcommand{\hamiltonian}{H}
\newcommand{\hilbert}{\mathscr{H}}
\newcommand{\perturbation}{\delta}
\newcommand{\unitary}{U}

\newcommand{\pauli}{\sigma}
\newcommand{\dt}{\ensuremath{\Delta_3}}

\begin{document}

\ifdefined\EMBEDDEDINRESPONSE
\else
\title{Integrable, Mixed, and Chaotic Dynamics in a Single All-to-All Ising Spin Model}
\author{
David Amaro-Alcal\'a$^{1,2,3}$\thanks{Corresponding author: \href{mailto:david.amaroalcala@savba.sk}{david.amaroalcala@savba.sk}}
\and
Carlos Pineda$^{4,5}$\thanks{\href{mailto:carlospgmat03@gmail.com}{carlospgmat03@gmail.com}}
}
\date{\today}

\maketitle
\begin{center}
\small
$^{1}$ Research Centre for Quantum Information, Institute of Physics, Slovak Academy of Sciences, D\'ubravsk\'a cesta 9, Bratislava 845 11, Slovakia\\
$^{2}$ Institute for Quantum Science and Technology, University of Calgary, Alberta T2N 1N4, Canada\\
$^{3}$ Department of Physics, Lakehead University, Thunder Bay, ON, P7B 5E1\\
$^{4}$ Instituto de F\'isica, Universidad Nacional Aut\'onoma de M\'exico, Ciudad de M\'exico 01000, Mexico\\
$^{5}$ Vienna Center for Quantum Science and Technology, Atominstitut, TU Wien, 1020 Vienna, Austria
\end{center}
\begin{abstract}
We demonstrate that the Ising \gls{ata} model exhibits a range of dynamics, from
integrable to chaotic, including mixed behaviour across symmetry blocks
within a single system. While other works have explored the dynamics of
all-to-all systems by varying parameters, we analyse a fixed set of parameters
and examine the dynamics within different blocks. In addition to investigating
the dynamical properties, we show that the system remains resilient to noise
when the norm of the Hamiltonian representing the noise is close to 1. Our
results are presented by mapping each symmetry sector of the system to a \gls{kt} and
observing that \gls{kt} parameters for each sector depend on its dimension.
This system, similar to the Bunimovich billiard for classical chaos, provides a
new platform for studying dynamics determined by the symmetry sector, advancing
quantum chaos research.
\end{abstract}
\fi

\section{Introduction}\label{sec:intro} 

Classical dynamical systems encompass a continuum of dynamics, ranging from
completely integrable to strongly chaotic~\cite{Kolmogorov1954,Arnold1963}. Integrable and chaotic regimes
are distinguished by stability, measured using the Lyapunov exponent, which
quantifies the divergence of trajectories in phase space~\cite{ott2002}. In chaotic
systems, trajectories traverse diverse regions of phase
space~\cite{lichtenberg1992}. Another
distinction is that integrable systems do not satisfy the \gls{eth} due to their
constraints, whereas chaotic systems do~\cite{srednicki1994,rigol2008}.

In quantum systems, dynamics are identified using statistical methods~\cite{HaakeBook,mehta2004random}. With
symmetry, it is standard to examine each invariant subspace independently~\cite{sakurai1994}.
Correlations typical of chaotic systems do not persist across subspaces~\cite{Pineda2014}.
Alternatively, the Krylov subspace approach divides the Hilbert space into
subspaces defined by evolution under a specific Hamiltonian~\cite{nandy2025}.

Different dynamics can coexist in the same system. Mixed dynamics feature
regions with chaotic behaviour and regions with integrable dynamics~\cite{Ozorio_1989}. Mushroom
billiards, for example, sharply separate integrable and chaotic regions based on
initial conditions~\cite{bunimovich2001,karel}. Unlike studies focusing on only two regimes, we show the
system displays a continuous range of dynamics across symmetry sectors. We
demonstrate that an \gls{ata} Ising spin system~\cite{lipkin1965} shows dynamics from fully
integrable to fully chaotic as the symmetry sector varies.

This analysis is done by noting that each symmetry block is a \gls{kt} and
studying the corresponding dynamics through the statistical properties of its
spectrum. In particular, we use the statistic $r$ (the \gls{nns}
ratio)~\cite{Oganesyan2007}, the \gls{nns}
distribution~\cite{berry1977,bohigas1984}, and the spectral rigidity
(Dyson–Mehta $\Delta_3$ statistic)~\cite{Haake1987}. With these statistics, we
affirm that the system exhibits not only integrable and chaotic statistics
but also mixed statistics~\cite{berryrobnik1984}. However, noise can affect the
determination of the properties of the
dynamic~\cite{Nonnenmacher2003,munozarias2021,Yoshimura2024,Li2025,Ferrari2025,Passarelli2025}.
Foreseeing an experimental setup, we analyse the effect of two types of noise (a
randomly sampled Hamiltonian and a spin chain with normally distributed weights
for each pairwise interaction). We show that until the norm of the Hamiltonian
perturbation is close to 1, the statistics can still be distinguished.

This paper is organised as follows. Section 2 covers the symmetries used to
analyse different systems, the physical systems discussed in the data, and those
involved in the perturbation process. It also outlines the \gls{rmt} tools employed to
examine the evolution. In Section 3, we detail the explicit decomposition and
the perturbation applied to assess the resilience of symmetries under noise.
Section 4 presents the results, showing, using statistical methods, that the \gls{ata}
spin system exhibits different dynamics depending on the symmetry sector. We
also employ the \gls{rmt} tools introduced in Section 2 to demonstrate that the
system's dynamics remain resilient even when subjected to a perturbation with a
norm of order~1. Finally, section~5 offers our conclusions.

\section{Background} 

In this subsection, we explain the symmetry reduction task we perform, the physical systems we use, the \gls{rmt} tools employed, and the methods to understand the subsequent sections, including our approaches and results. In the first section, we introduce the concept of symmetry decomposition, which enables the study of large \gls{ata} spin systems. In Subsection~\ref{sec:physical-system},
we discuss the physical systems we employ. Lastly, in Subsection~\ref{sec:intro-rmt}, we outline the three statistical measures used to characterise our studies on chaos.

\subsection{Symmetry sectors}\label{sec:symmetry-sectors} 

In this subsection, we recall the concept of the symmetry sector. The concept of symmetry is used as a form of invariance under the action of a group.
In the present work,
the physical system consists of a finite number of spin ½ particles. Therefore, the relevant group is U(2), and since global phases are irrelevant in our study, we concentrate on SU(2).

We recall the notation of a symmetry sector. Formally, a symmetry sector in our case is then defined as an invariant subspace of the total Hilbert space under the action of SU(2). In other words, a decomposition of the Hilbert space into irreducible representations (irreps) of SU(2).
This is a well-known topic, so we should simply recall its main notion and the notation relevant to the following sections.

Now we discuss the Hilbert space decomposition into symmetry blocks.
Let~\(\gamma\) be the fundamental irrep of SU(2) and also the Hilbert space of a single spin ½ particle.
For \(\spinnumber\) spins, the Hilbert space is
\begin{equation}
\hilbert \coloneqq
\gamma^{\otimes \spinnumber} .
\end{equation}
This is equivalent to the Hilbert space of \(\spinnumber\) spins ½.
Using this fact,
we denote the decomposition of $\hilbert$ into irreps as
\begin{equation}\label{eq:invariance}
  \hilbert \cong \bigoplus_j \Gamma_j^{\oplus  m_j} ,
\end{equation}
where the sum symbol denotes direct sum, and \(m_j\) denotes the multiplicity of each spin~\(j\) in the decomposition.
Each \(\Gamma_j\) labels a subspace of size \(2j+1\) invariant under the irrep labelled by  \(j\) of SU(2).
This decomposition can be obtained in many ways---one using Clebsch–Gordan
coefficients, which we do not explain here.

The states of \(\hilbert\) are denoted \(\ket{J,M}\); that is, \(\hilbert =
\langle \ket{J,M}\rangle\). In turn, the states for each~\(\Gamma_j\) are
denoted \(\ket{j,m}\). Whereas some \(\Gamma_j\) appear more than once, we are
only interested in a specific block, thus we ignore the multiplicity in the
notation.

\subsection{Physical systems}\label{sec:physical-system} 

In this subsection, we introduce the three physical systems we use and their
parameters. Two are important for the dynamics of \gls{ata} kicked systems,
and the last is
used for the perturbation study. For the \gls{kt} system, we also show the link
between the parameters of the classical \gls{kt} with integrable and chaotic
dynamics.

We begin by defining the Hamiltonian for the \gls{ata} system and also for the magnetic interaction with the spin system.
In the \gls{ata} coupling scenario, the interaction becomes
\begin{equation}\label{eq:ising-all-to-all}
  H_\subata(\tauata) \coloneqq \tauata\sum_{i<k} \sigma_i^z \sigma_k^z.
\end{equation}

\label{par:meanfield}

Because the coupling is \gls{ata}, the interaction can be written entirely in terms of collective spin operators (e.g., \(J_z=\frac12\sum_i \sigma_i^z\) and
\(J_z^2\)). In the large-\(N\) limit, the collective variables behave classically,
so the mean-field (semiclassical) dynamics is precisely the classical \gls{kt} map given in Eq.~\eqref{eq:stroboscopic-evolution}~\cite{carollo2021}. The kicked-top correspondence therefore makes the mean-field reduction explicit, while remaining exact at finite \(N\) within each symmetry block. In particular, the all-to-all Hamiltonian is a standard mean-field quantum spin model: it depends only on collective spins and preserves the symmetric subspace, so each symmetry sector is a single large spin~\cite{Bapst2012}.

The kick Hamiltonian is
\begin{equation}\label{eq:kick-chain}
H_\subkick(b_x) \coloneqq b_x\sum_{i}  \sigma_i^x,
\end{equation}
where \(\sigma_i^x\) is the $x$ Pauli matrices acting on the \(i\)-th spin, and \(b_x\) represents a constant magnetic field.

The kicked system is then described by the Hamiltonian
\begin{equation}\label{eq:hamiltonian-kicked-all-ata}
  H_{\subkick\subata}(t;b_x, \tau_\subata ) \coloneqq
 H_\subkick(b_x) \sum_{n\in \mathbb{Z}} \delta(t-n)+  H_\subata(\tauata) .
\end{equation}

From Eq.~\eqref{eq:hamiltonian-kicked-all-ata},
we define the unitary evolution of the  \gls{ata} system. The unitary evolution of the system combines the contribution of the Ising chain or
\gls{ata}, followed by the periodic kick.
The periodical (also called stroboscopic) evolution is given by the Floquet operator~\cite{Stckmann1999}. Assuming homogeneous parameters, the Floquet operator for the \gls{ata} interacting system is given by
\begin{equation}\label{eq:ata}
  U_{\subkick\subata}(\tau_\subata,b_x) \coloneqq
  U_\subkick(b_x)
  U_{\subata}(\tau_\subata),
\end{equation}
with
\begin{equation}\label{eq:ising-term-all}
  U_{\subata}(\tauata) \coloneqq
  \exp\left(-\myi  H_\subata(\tau_\subata)\right) \text{ and }  U_{\subkick}(b_x)
\coloneqq \exp\left(-\myi H_\subkick(b_x)\right).
\end{equation}

We now recall the evolution, both classical and quantum, of the \gls{kt} system.  We use this to introduce notation, the parameters that exhibit integrable and chaotic dynamics, and the evolution operator.

The classical \gls{kt} is a system characterised solely by angular momentum. The \gls{kt} Hamiltonian is
\begin{equation}
  H(\alpha, \tau_\subtop) = \hamZ(\alpha)\sum_n \delta(t-n) + \hamO(\tau_\subtop) ,
\end{equation}
with
\begin{subequations}
\begin{align}
  \hamZ(\alpha) &\coloneqq \alpha J_x, \\
  \hamO(\tau_\subtop) &\coloneqq \taukt J_z^2;\label{eq:evolution-h-1}
\end{align}
\end{subequations}

where \(\hamZ\) generates a rotation about the x-axis with angular velocity
\(\alpha\), and \(\hamO\) implements the nonlinear twist proportional to \(J_z^2\).

We now discuss the evolution of the \gls{kt}. We use Hamilton equations to obtain the evolution of the system. The evolution is first given in terms of the spin system (either chain or \gls{ata}) and then the kick. The stroboscopic evolution of the system is given by the mapping
\begin{equation}\label{eq:stroboscopic-evolution}
F\colon ( J_x^{(n)}, J_y^{(n)}, J_z^{(n)}) \to ( J_x^{(n+1)}, J_y^{(n+1)}, J_z^{(n+1)}),
\end{equation}
with
\begin{subequations}
  \begin{align}
    J_x^{(n+1)} &=
    J_{x}^{(n)} \cos \taukt \bigl(J_{y}^{(n)} \sin \alpha + J_{z}^{(n)} \cos \alpha \bigr) \\
        &\quad - \bigl(J_{y}^{(n)} \cos \alpha - J_{z}^{(n)} \sin \alpha \bigr)
        \sin \taukt \bigl(J_{y}^{(n)} \sin \alpha + J_{z}^{(n)} \cos \alpha \bigr),
        \nonumber\\[6pt]
    J_y^{(n+1)} &=
    J_{x}^{(n)} \sin \taukt \bigl(J_{y}^{(n)} \sin \alpha + J_{z}^{(n)} \cos \alpha \bigr) \\
        &\quad + \bigl(J_{y}^{(n)} \cos \alpha - J_{z}^{(n)} \sin \alpha \bigr)
        \cos \taukt \bigl(J_{y}^{(n)} \sin \alpha + J_{z}^{(n)} \cos \alpha \bigr),\nonumber \\[6pt]
    J_z^{(n+1)} &=
    J_{y}^{(n)} \sin \alpha + J_{z}^{(n)} \cos \alpha .
  \end{align}
\end{subequations}
The resulting vector is also normalised.

\begin{figure} 
  \centering
  \includegraphics[width=0.7\textwidth]{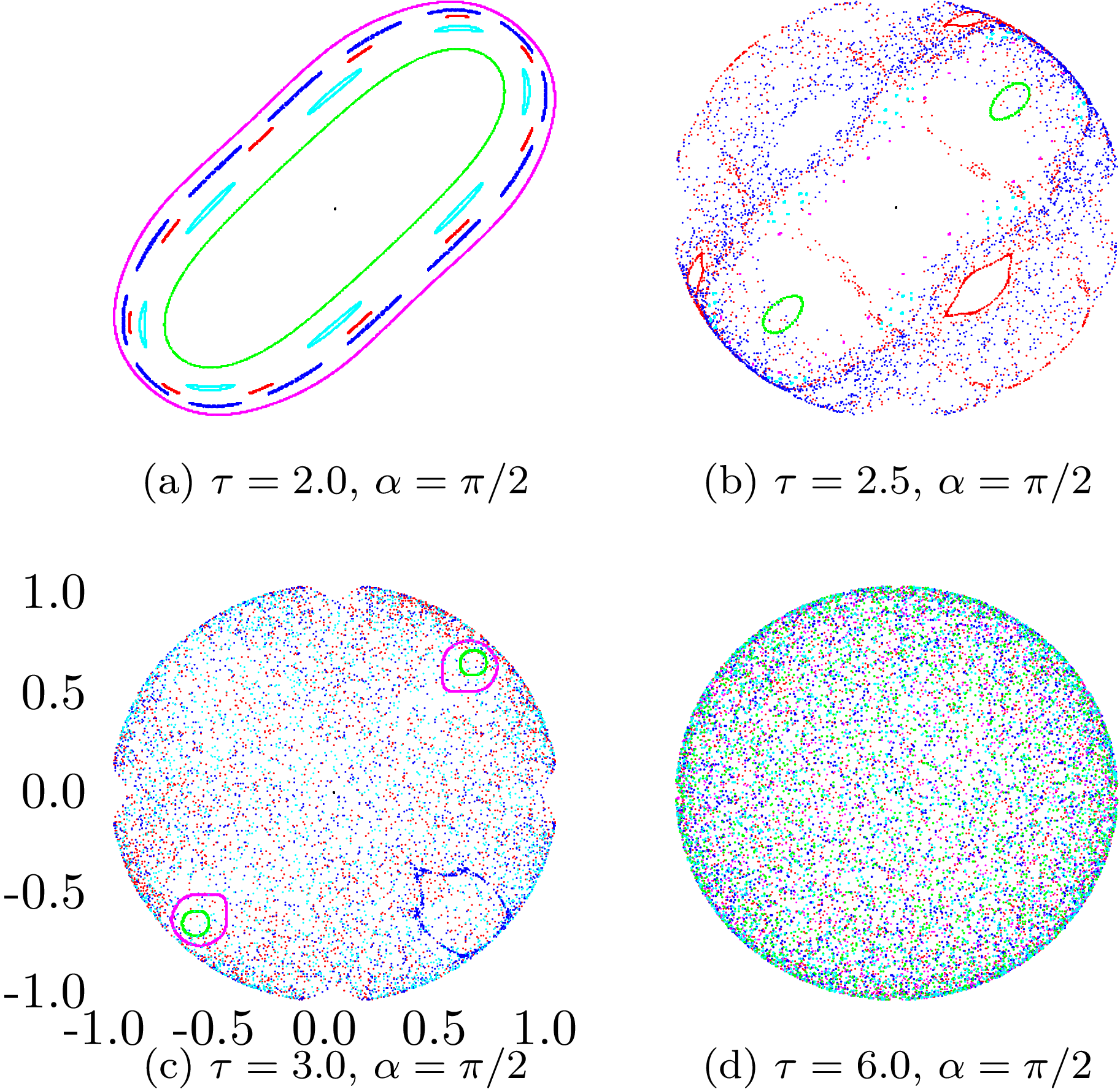}
\caption{\label{fig:phase-space-classical} Phase space, \(J_x\) \emph{vs} \(J_z\), for the evolution of the \gls{kt} with the parameters tau and alpha mentioned. The range of the variables is the same as the frames in plot c). The colours in the plot correspond (up to normalisation) to different initial conditions \((J_{x}^{(n)},J_{y}^{(n)},J_{z}^{(n)}) \): red corresponds to \((-1,-3,-3)\), blue \((-1, -13/10, -2)\), magenta \((-1, -0.3, -1)\), green \((1, -1, 1)\), cyan \((0, -2, 1)\), and black \((0, -2, 2,0)\). }
\end{figure} 

We now describe how we use the mapping \(F\) to obtain pictures of the phase
space. Applying \(F\) on a unitary vector~\( ( J_x^{(0)}, J_y^{(0)}, J_z^{(0)})
\), one obtains the  slices of the phase space of the classical \gls{kt} using
the parameters mentioned in each image, Figure~\ref{fig:phase-space-classical}.
In these images, one can observe the disappearance of islands of regular motion
and the gradual increase of chaotic dynamics, all this as the parameter \(\tau\) increases.

Now we describe the Floquet operator for the \gls{kt}. To connect with the \gls{kt} system, we recall the definition of the collective spin operators:
\begin{equation}\label{eq:spin-angular-momentum}
  J_q \coloneqq \frac{1}{2}\sum_i \sigma^{q}_i,
\quad q \in \{x, y, z\},
\end{equation}
and define the total angular momentum squared as
\begin{equation}
  \mathbf{J}\cdot\mathbf{J} \coloneqq J_{x}^2 + J_{y}^2 + J_{z}^2 = \spinsize^2.
\end{equation}
The Floquet operator, in natural units, for the \gls{kt} system, is then given by
\begin{equation}\label{eq:evolution-kt}
  U_\text{KT}(\taukt,\alpha) \coloneqq
\exp \left(-\myi \alpha J_x\right)
  \exp \left(-\myi\frac{ \taukt J_z^2}{2\spinsize + 1}\right),
\end{equation}
describing stroboscopic evolution under alternating quadratic and linear spin terms.

We recall the definition of a spin chain and the kick and then introduce the \gls{ata} system. For nearest-neighbour interactions, the Ising Hamiltonian is
\begin{equation}\label{eq:ising-chain}
  H_\subchain \coloneqq \sum_{i=0}^{\spinnumber-1}
\sigma_i^z \sigma_{i+1}^z,
\end{equation}

where \(\sigma_j^z\) denotes the Pauli \(z\)-matrix acting on spin \(i\).

Similarly, the Floquet operator for the nearest-neighbour model is
\begin{equation}
U_{\subkick\subchain}(\isingweight, b_x)
=
U_{\subkick}(b_x)
U_{\subchain}(\isingweight),
\end{equation}
with
\begin{equation}\label{eq:ising-term}
  U_{\subchain}(\isingweight)
\coloneqq \exp\left(-\myi \isingweight H_\subchain\right).
\end{equation}

We finish this section by discussing the notation for operators invariant in symmetry subspaces. Consider a unitary operator \(U\) that is invariant in each irrep \(\Gamma_j\) in Eq.~\eqref{eq:invariance}. We thus denote the restriction of \(U\) onto \(\Gamma_j\) as \(U^{(j)}\). Therefore, we have the following equivalence
\begin{equation}\label{eq:block-decomposition}
U \cong
\bigoplus_j \left(U^{(j)}\right)^{\oplus  m_j}.
\end{equation}

for simplicity, if from the context it is clear that we are dealing with a
single block in Eq.~\eqref{eq:block-decomposition}, then we omit the superscript \((j)\).

\subsection{Random matrix theory tools}\label{sec:intro-rmt} 

Studying the evolution of periodically driven systems, instead of energies, we
use the quasienergy spectrum. These are defined as the eigenphases of the
Floquet operator~\cite{Stckmann1999}. The statistical properties of the
eigenphases allow distinguishing Poissonian or \gls{rmt}
statistics~\cite{Haake1987,HaakeBook}. To assess the chaotic dynamic of the
system, we use standard statistical measures from random matrix theory.  These
include both local and long-range spectral statistics, applied to ideal and
noisy evolutions.  We use the long-range statistics to confirm the results
provided by the local statistics.

\begin{figure}[ht] 
  \centering
  \includegraphics{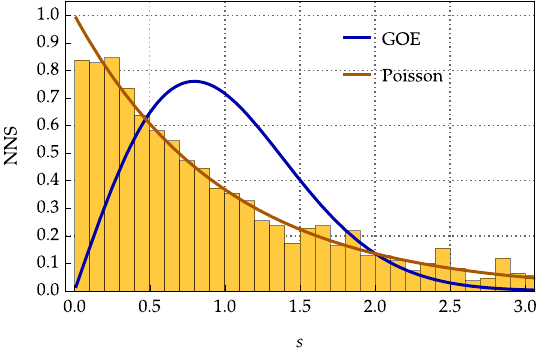}
\caption{\label{fig0} Spacing distribution for \(N = 101\), \(J = 801\), \(\alpha = 1.7\), and \(\tau \in [10,10.5]\) with step size \(0.001\).  The horizontal axis shows normalised spacings, while the vertical axis shows the estimated probability density.}
\end{figure}

\begin{figure}[ht] 
  \centering
  \includegraphics{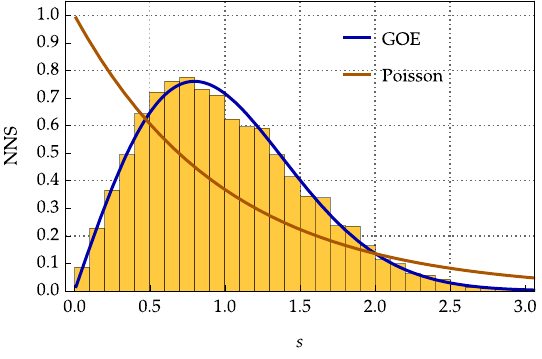}
\caption{\label{fig1} Spacing distribution for \(N = 701\), with other parameters identical to Figure~\ref{fig0}. }
\end{figure}

We now recall the definitions of the \gls{nns} distribution and the statistic
\(r\), which we use to study short-range correlations. The \gls{nns}
distribution is the statistic resulting from obtaining the absolute value of the
difference between adjacent quasienergies. We compare the empirical distribution
of eigenphase spacings with theoretical distributions from \gls{rmt} and Poisson
ensembles, which serve as indicators of chaotic and integrable dynamics,
respectively.
 We present examples for the Wigner
and Poisson in Figs.~\ref{fig0} and~\ref{fig1}.
The~\(\tilde{r}\)-statistic provides a robust alternative to the \gls{nns} distribution without requiring spectral unfolding~\cite{Oganesyan2007,Atas2013}. It is defined as
\begin{equation}\label{eq:definition_statistic_r}
  \tilde{r}_n \coloneqq
  \frac{\min(s_n, s_{n-1})}{\max(s_n, s_{n-1})},
\end{equation}
where \(s_n\) is the \(n\)-th eigenphase spacing. This dimensionless quantity is bounded in~\([0, 1]\), with known ensemble averages:
\begin{itemize}
\item Poisson: \( r = 2 \ln 2 - 1 \approx 0.386\),
\item \gls{goe}: \(r = 4 - 2 \sqrt{3} \approx 0.536\),
\end{itemize}
where $r \coloneqq \langle\tilde{r} \rangle$ is the average over \(n\) (over each spacing) in Eq.~\eqref{eq:definition_statistic_r}.

We now discuss another statistic that serves to study long-range correlation in
the eigenphases. The \(\Delta_3\) statistic~\cite{Mehta1963} quantifies
long-range correlations in the spectrum and is more sensitive to chaotic
dynamics than local statistics. This statistic is computed as follows. Given an ordered set of eigenphases \(S = \{s_1, s_2, \dots, s_\spinnumber\}\), we define the staircase function \(N(s)\) as the number of eigenvalues less than or equal to \(s\). We then fit a linear function \(f(s) = a s + b\) to \(N(s)\) and calculate the deviation. The \(\dt\) statistic is obtained by integrating this deviation:

\begin{equation}
  \Delta_3(L) \coloneqq
\int_0^{L} \left[f(s) - N(s)\right]^2 \, \mathrm{d}s,
\end{equation}
where \(L\) is the maximum spacing.

In this section, we outline our methods for studying the \gls{ata} system. We first discussed the physical system and some other relevant aspects for the study. Using the block symmetries of the \gls{ata} system, we demonstrate that a
\gls{kt} is obtained. We then reviewed three statistics we use to study chaos
signatures in the \gls{ata} system.

\section{Approach} 

In this section, we describe the tool application to reach our objectives. We discuss the parameters of the classical model and how we can obtain a transition between the models.  Then, we introduce two kinds of perturbations that we use in Section~\ref{sec:results} to show the robustness of the effect.

\subsection{Symmetry sector decomposition of the ATA} \label{sec:ata-is-kt-and-kt-is-ata} 

In this subsection, we present two key arguments necessary for our study of the dynamics of the \gls{ata} system. First, we demonstrate that the \gls{ata} evolution commutes with the total angular momentum squared, enabling the evolution operator to be block-diagonalised. We argue that each block corresponds to a \gls{kt} system. Finally, we conclude the subsection by mapping the parameters from the
\gls{kt} to the \gls{ata} system.

Our proof of the equivalence between the \gls{ata} system and the \gls{kt} begins by showing that \(H_\subata\) is a function of \(J_z\). Recalling the definition of \(J_z\) of Eq.~\eqref{eq:spin-angular-momentum}, we note that:

\begin{equation}
J_z^2 = (\sigma_1^z + \cdots + \sigma_\spinnumber^z)(\sigma_1^z + \cdots +
\sigma_\spinnumber^z) = \frac{1}{2\tauata}H_\subata(\tauata) + \frac{1}{4}\spinnumber\mathbb{1},
\end{equation}
or equivalently
\(H_\subata(\tauata) = 2\tauata\,(J_z^2 - \frac{1}{4}\spinnumber\mathbb{1})\).
Also, $H_{\subkick}$ is proportional to $J_{x}$. Thus, the evolution operator of
the \gls{ata} system commutes with \(\mathbf{J}\cdot \mathbf{J}\).  Therefore,
we individually study \(U_{\subkick\subata}\) on each \(\Gamma_j\) separately.

Our next goal is to justify that the restriction of \(U_{\subkick\subata}\) into \(\Gamma_j\) is a \gls{kt}. Writing~\(H_\subata\) and \(H_\subkick\) in terms of $J_{z}$ and $J_{x}$ reveals that
\begin{equation}\label{eq:equivalence-kt-ata}
 \exp(-\myi  H_{\subkick}(\alphaata)) \exp(-\myi  H_{\subata}(\tauata)) \\=
\exp(-\myi \alphaata J_{x})
\exp\left( -\myi \frac{\tauata N}{2} \right)
\exp\left( -\myi 2\tauata J_{z}^{2} \right).
\end{equation}
Since the restriction of the operators \(J_z\) and \(J_x\) acting on
\(\hilbert\) is again a spin operator~\(J_z\) and~\(J_x\) acting now on
\(\Gamma_j\), each block with fixed angular momentum $j$ of the \gls{ata} system
is equivalent to a \gls{kt}. The global phase (the first term in the
multiplication) is ignored, and the operators $J_{x}$ and $J_{z}$ are block
diagonalised, leading to the corresponding spin operator for $j$. From now on,
the operators \(J_x\) and \(J_z\) refer to the \(j\)-spin operators acting on
\(\Gamma_j\) and not to the ones acting on \(\hilbert\).

To conclude this section, we relate the parameters of the \gls{ata} model and
the \gls{kt}. From Eq.~\eqref{eq:equivalence-kt-ata}, we can see that the
symmetry sector $j$ of the \gls{ata} model has exactly the same Hamiltonian as
the \gls{kt} if we identify

\begin{equation}\label{eq:transformation-parameter}
  \alpha = b_x\quad \text{and}\quad
  \taukt = \tauata\frac{j}{2(2\spinsize+1)}.
\end{equation}

This means that the \gls{ata} model is equivalent to a direct sum of \gls{kt}s,
each with dimension~\(j\), and each with a varying set of parameters. In
particular, notice the dependence of \(\taukt\) on~\(j\). This parameter, in the
\gls{kt}, can be varied to control the degree of chaoticity in the system
\cite{Haake2010}. This implies that a carefully tuned \gls{ata} model carries the dynamics of various \gls{kt}s. This is exploited in \sref{sec:results}.

\subsection{Perturbation}\label{sec:perturbation} 

In this subsection, we discuss two physically motivated perturbations that we later use to study the effect of noise on the statistics. We study two types of perturbation, one is an issue of noise addressing individual pairs of spins in the \gls{ata} system, and the other represents arbitrary noise in the Hamiltonian.

The perturbations, thus, do not affect the kick part of the evolution.
Therefore, the perturbations we use correspond to a contribution to the
Hamiltonian \(\hamiltonian_\text{T}\) multiplied by a perturbative factor.

Here, we justify why we can study the global perturbation in a single block. We
start with a global \gls{goe} perturbation. The spectrum of a block diagonal
\gls{goe} matrix (a matrix which is a direct sum of \gls{goe} members) is the
same as that of a \gls{goe} matrix with the same dimension. For the case of an Ising chain perturbation, the action commutes simply because both are diagonal in the computational basis.

We also add a constraint to \(\delta\).
For a general perturbation Hamiltonian \(H_\text{pert}(\delta)\), where~\(\delta \ll 1\) is chosen so that,
\begin{equation}
  [H_\subata(\tauata, b_x), H_\text{P}(\delta)] \sim 0;
\end{equation}
ensuring \(\delta\) is sufficiently small for the perturbation to almost commute with the
original evolution.

We impose \(\delta\ll 1\) to have the form of the perturbation simply as the product of the unitary perturbation times the ideal evolution.

We now describe how the perturbation is introduced into the system. The noise is
added to the Hamiltonian in Eq.~\eqref{eq:ising-all-to-all} in the form~\(\delta
H'\), where \(H'\) is the perturbation and~\(\delta\ll 1\). Thus, we have
\begin{equation}\label{eq:deltaHamiltonian}
\exp(\myi (\hamO(\taukt) + \delta H'))
\approx
\exp(\myi \hamO(\taukt) )
\exp(\myi  \delta H'),
\end{equation}
with \(\hamO\) introduced in Eq.~\eqref{eq:evolution-h-1}. Our perturbations are added to the Ising term of the stroboscopic evolution.

The first type of perturbation we study is a random Hamiltonian.
We achieve this perturbation by adding a \gls{goe} matrix~\cite{mehta2004random} to the Hamiltonian.
The \gls{goe} perturbation is applied as
\begin{equation}
  \tilde{H}_{\mathrm{T}}(\taukt) \coloneqq  \hamO(\taukt) + \delta H_{\mathrm{GOE}},
\end{equation}
where \(H_{\mathrm{GOE}}\) is a randomly sampled \gls{goe} matrix with dimension
\(2j+1\).

Because~\(\delta\ll 1\), the noisy stroboscopic evolution is
\begin{equation}\label{eq:goe_perturbation_evolution}
  \tilde{U} =
\exp\left(-\myi \alpha J_x\right)
  \exp(-\myi \delta H_{\mathrm{GOE}} )
  \exp\left(-\myi \frac{\taukt}{2(2\spinsize+1)} J_z^2\right).
\end{equation}

The second perturbation is an Ising chain with normally distributed weights. This perturbation is chosen to simulate errors in addressing individual spins. The Hamiltonian is
\begin{equation}\label{eq:hamiltonian-perturbation-ising}
H_{\text{R}\subchain}(\boldsymbol{\delta})
\coloneqq
\sum_i \delta_i \pauli_i \pauli_{i+1},
\end{equation}
with \(\delta_i \sim \normaldist(0,\delta^2)\). Here,
\(\boldsymbol{\delta}\coloneqq (\delta_0,\ldots,\delta_{\spinnumber-1})\).

Besides the weights \(\delta_i\), we use \(\delta\) to model the
perturbation's strength.
This concludes our exposition of the types of perturbations we use to study the
system's resilience.

In this section, we covered three points. Firstly, we showed how the evolution operator can be decomposed to act on a spin-\(j\) subspace. Next, we explained that the restriction on each block is equivalent to a \gls{kt}. Lastly, we outlined the two types of perturbation we will use in Sec.~\ref{sec:results} in the context of our findings.

\section{Results}\label{sec:results} 

In this section, we present two main results. First, we demonstrate that a single physical system with a fixed set of parameters—the \gls{ata} model—can display a spectrum of dynamics, from integrable to chaotic, depending on the symmetry block considered. We characterise these regimes using the \(r\) statistic, the \gls{nns} distribution, and~\dt, covering a broad range of energies. Second, we quantify the robustness of these dynamical features against noise. By introducing two physically motivated perturbations, we show that the dynamics in all symmetry blocks converge to a common statistical dynamic: either Poissonian or \gls{rmt}, depending on the nature of the perturbation. Notably, this transition occurs when the norm of the perturbation approaches unity.

\subsection{Chaotic signatures across different symmetry blocks} 

In this subsection, we analyse the statistics by symmetry blocks of the \gls{ata} system. We do so to assess its level of chaos or integrability. To carry out this study, we separate the spectra of the \gls{ata} system into symmetry sectors. As discussed in Sec.~\ref{sec:ata-is-kt-and-kt-is-ata}, each symmetry block of the \gls{ata} system is equivalent to a \gls{kt}. Each block's \gls{kt} shares the same global parameters as the original \gls{ata} system. However, we observe different statistics across blocks due to their varying sizes. We now detail the parameters further.

We now list the parameters we use to carry out our study. Furthermore, we fix
the dimension of the \gls{ata} system to~\(\spinsize = 801\); that is, an
\gls{ata} system with \(\spinnumber=400\) spins ½. The parameters for the
\gls{ata} system are \(\alpha = 1.7\) and \(\taukt\) within the range \([10,
10.5]\), considering~\(501\) equally spaced values; that is, we
consider~\(\taukt =10.0, 10.001, 10.002, \ldots, 10.5\). The selection of these
parameters is explained in Sec.~\ref{sec:physical-system}, and is motivated by
the classical \gls{kt}; the classical \gls{kt} exhibits classical chaos for
those parameters~\cite{Haake2010}.

\begin{figure} 
  \centering
  \includegraphics[]{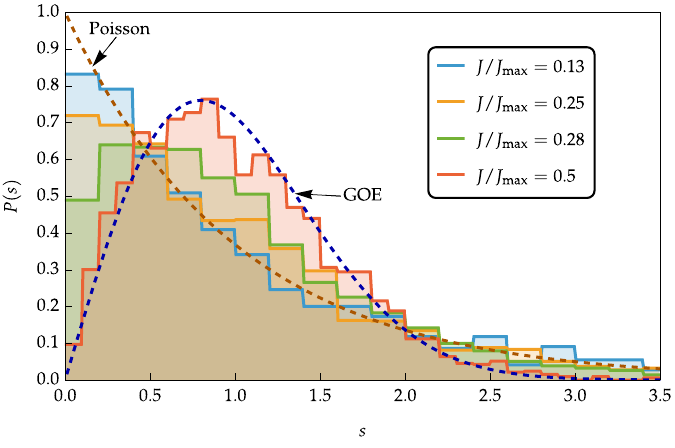}
\caption{ \label{fig:NNS distribution-ideal-different-sizes-transition} \gls{nns} distribution for four sets of block sizes with a system with the same configuration. The dimensions of the irreps considered are $\spinsize = 101, 197, 225$, and $401$, with~$\spinsize_{\max} = 801$. The blue dashed line represents the \gls{nns} distribution for \gls{goe}, and the orange dashed line indicates the Poisson \gls{nns} distribution. }
\end{figure}

We analyse our results, showing intermediate statistics that differ from both the Poisson and \gls{goe} distributions, appearing in various symmetry blocks. Figure~\ref{fig:NNS distribution-ideal-different-sizes-transition} illustrates the \gls{nns} distribution for two block sizes, highlighting differences between blocks. Using the statistic \(r\), defined in Eq.~\eqref{eq:definition_statistic_r}, across symmetry blocks, Figure~\ref{fig:delta-three-comparison} demonstrates a transition in \(r\) values from Poisson to \gls{goe} characteristics\label{par:poisson-goe}. Each block has a distinct value until it converges to the \gls{goe}, at which point it stops changing; equivalently, increasing block dimension drives a smooth interpolation from Poisson to \gls{goe} statistics.

Physically, this transition does not require changing the Hamiltonian: it reflects which symmetry sector the initial state populates, with different \(J\) blocks displaying regular (Poisson) or chaotic (GOE) statistics.

\begin{figure} 
  \centering
  \includegraphics[]{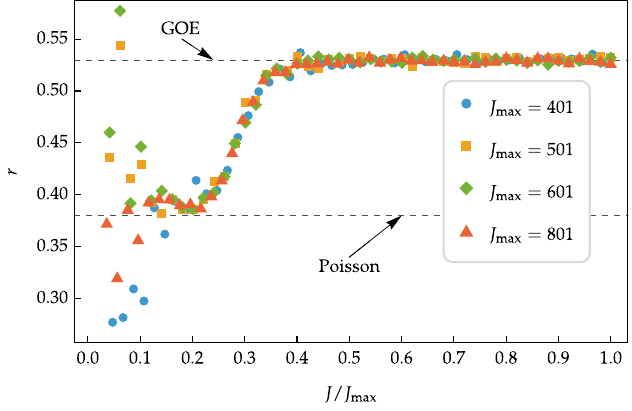}
\caption{\label{fig:comparison-stat-r-blocks} Value of the statistic $r$ relative to the block size. The block size is divided by the total block size. The grey dashed lines show the \(r\) values for Poisson and \gls{goe}. }
\end{figure} 

 Figure~\ref{fig:comparison-stat-r-blocks} was computed by using different systems with the same parameters. Then, for each system, different symmetry blocks were studied. Consider, for instance, the blue points. These correspond to a system with \(\spinnumber = 802\) spins \(1/2\). We then study different blocks for such a system; we chose, for convenience, the odd-dimensional blocks. We compute the statistic \(r\) for those blocks and plot the quotient
\(\spinsize/\spinsize_{\max}\) (for the blue points,~\(\spinsize_{\max}=401\))
versus the statistic \(r\) for the block of dimension \(\spinsize\). We do that
for~\(\spinsize_{\max} = 401, 501, 601,\) and \(801\). This shows that for any
number of spin \(1/2\), we always get the same curve for
\(\spinsize/\spinsize_{\max}\) versus \(r\).

We now discuss our results for the statistics \(\dt\). The \gls{nns} distribution method offers a straightforward yet limited way for assessing statistics. Since it is a local measure, there may be cases where the \gls{nns} distribution indicates either
\gls{goe} or Poisson, but the overall dynamic can differ.
Therefore, we use the~\(\dt\) statistic to evaluate the long-range correlations. We analyse the \(\dt\) using the same parameters as in Figure~\ref{fig:delta-three-comparison}, with values of 101 and 301. The results in Figure~\ref{fig:delta-three-comparison} show that the long-range dynamic matches the expected statistics: Poisson for \(\spinsize = 101\) and \gls{goe} for~\(\spinsize = 301\). These findings confirm that the \gls{ata} system exhibits both Poisson and \gls{goe} statistics in different symmetry sectors.

\begin{figure}[ht] 
  \centering
\includegraphics{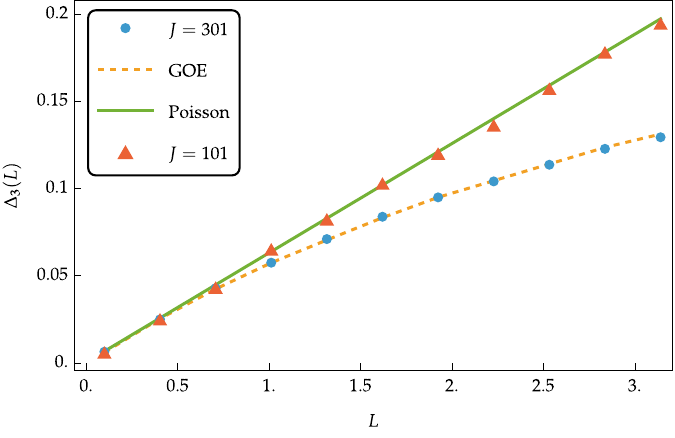}
\caption{\label{fig:delta-three-comparison} Plot spacing \(L\) against \(\Delta_3(L)\) for data sampled from \gls{goe} and Poisson ensembles, and numerical estimates from the \gls{ata} system. }
\end{figure} 

We now conclude this subsection. We have shown that a single system exhibits a
range of statistics from \gls{goe} to Poisson. The analysis employed both local
and global chaos indicators, namely the \gls{nns} distribution and \(\dt\) statistics. These markers agree in their characterisation, revealing two types of statistics and a transition between intermediate values. In the following subsection, we explore how perturbations affect this ideal dynamic.

\subsection{Robustness under perturbations} 

In this subsection, we examine the impact of two kinds of perturbations on the system.

Our focus is on assessing the resilience of the overall system under stress from these perturbations. The analysis primarily involves comparing the resulting statistic \(r\) with a measure of the perturbation strength.

Accordingly, the stroboscopic evolution of the perturbed system is given by

\begin{equation}
\tilde{U}_\subata(\alpha, \tau, \spinsize;\delta)
=
U_\text{P}(\delta)
U_\subata(\alpha, \tau, \spinsize).
\end{equation}
We now recall two kinds of unitary perturbations introduced in Sec.~\ref{sec:perturbation}. The first type of perturbation is generated from a \gls{goe} Hamiltonian, leading to the unitary operator
\begin{equation}
  \unitary_\text{GOE} \coloneqq \exp(-\myi \delta H_\text{GOE}).
\end{equation}
The second type corresponds to an Ising chain, where the coupling coefficients
\(\delta_i\), collected as
\(\boldsymbol{\delta}\coloneqq (\delta_0,\ldots,\delta_{\spinnumber-1})\), are randomly
sampled from the normal distribution \(\normaldist(0, \delta)\), resulting in the
unitary operator
\begin{equation}
  \unitary_{\text{R}\subchain}(\boldsymbol{\delta})
  \coloneqq
  \exp\left(-\myi \sum_i \delta_i \pauli^{z}_i \pauli^{z}_{i+1}\right).
\end{equation}
We now begin our analysis.

We again use the statistic \(r\) to study changes in the system's dynamics.
First, we compute the \gls{nns} distribution for the system as the strength of the perturbation increases. Then, for each value of \(\tau\) in the range \([10.0,10.5]\), we select a representative sample of the corresponding perturbation. In the case of the Ising chain, the coupling constants \(\delta_i\) are uniformly randomly sampled from the interval \([0, 2\pi]\). For the \gls{goe} case, a different realisation from the ensemble is used for each perturbation. Additionally, we keep track of the norm of the perturbation throughout the process.

\begin{figure} 
\centering
\includegraphics{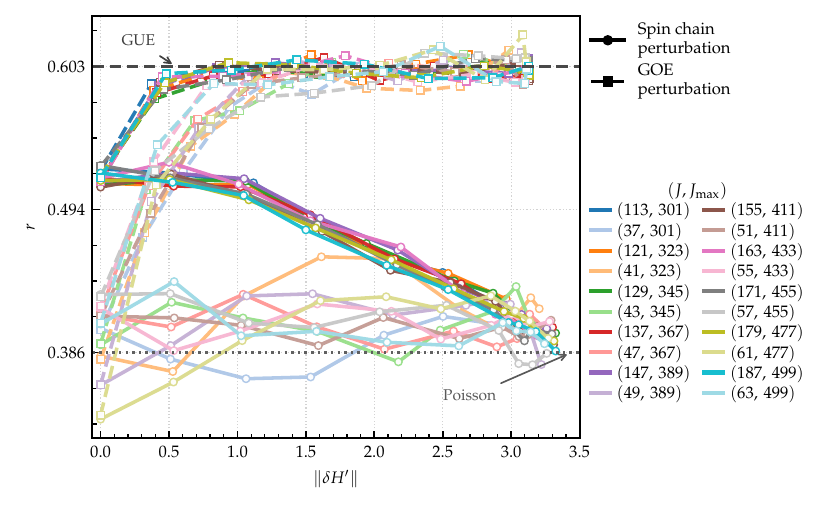}
\caption{\label{fig:comparison-stat-r-comprehensive} Plot of the level-spacing statistic \( r \) as a function of the perturbation norm \( \lVert \perturbation \hamiltonian^{\prime} \rVert \). We consider two sets of blocks with values of \( \spinsize / \spinsize_{\max} \approx 0.126 \) and \(0.376\), corresponding to integrable and chaotic dynamics, respectively. The perturbation consists either of the spin-chain Hamiltonian (circles) or a random GOE perturbation (squares), introduced in Eqs.~\eqref{eq:hamiltonian-perturbation-ising} and~\eqref{eq:goe_perturbation_evolution}. Different system sizes, characterised by different values of \( \spinsize_{\max} \), are included; in all cases, the curves tend to collapse onto a common profile. Notably, for sufficiently large perturbations, the value of \( r \) depends only on the nature of the perturbation. Moreover, the crossover occurs at approximately the same perturbation-norm value in both dynamical regimes and across all system sizes, which suggest a stable behaviour in the thermodynamical limit.}

\end{figure}

We present the corresponding results in Figure~\ref{fig:comparison-stat-r-comprehensive}.  There, we show the value of the statistic \(r\) for various values of the perturbation, quantified by the norm of \(\delta H'\) in Eq.~\eqref{eq:deltaHamiltonian}. In each case, the ideal system—that is, without perturbation—corresponds to the parameters used in the previous section to demonstrate that the \gls{ata} system can display a whole range of statistics, in particular \gls{goe} and Poisson.

In addition,
Figure~\ref{fig:comparison-stat-r-comprehensive} shows that increasing the perturbation causes the statistic \(r\) to shift from its original value to a different one. We consider two dynamical regimes
\( \spinsize / \spinsize_{\max} \approx 0.126\), \(0.376\)

corresponding to initial Poisson-like and \gls{goe}-like values respectively.

In the case of a \gls{goe}-type perturbation, the convergence occurs at the value of~\(r\) associated with the \gls{gue}:
\(2 \sqrt{3} \pi^{-1} - 2^{-1} \approx 0.60266\).
The convergence is seen as both the statistics for blocks corresponding to Poisson and \gls{goe}-like statistics decrease (or increase) to reach a value of the statistic \(r\) corresponding to \gls{gue}. In contrast, for the Ising chain perturbation, the convergence approaches a value close to that of the Poisson distribution, which is~\(2\ln 2 - 1 \approx 0.38629\).

By repeating the analysis for ten values of \(\spinsize\),
we also show that this behaviour is independent of the system size and that the
overall behaviour is the same for any size.

\section{Conclusions} 

Quantum and classical systems can display integrable, chaotic, or mixed dynamics. For quantum systems, statistical methods such as \(r\)-statistics,
\gls{nns} distribution, and rigidity are used to determine the kind of dynamics.
In the classical case, the divergence or convergence of initially close trajectories in the phase space is one way to determine the dynamics.

We classify the symmetry blocks by mapping each to an effective \gls{kt} model,
which provides a simple route to identifying parameter regimes associated with
either integrable or chaotic dynamics.
Furthermore, we demonstrate that different symmetry blocks can have mixed statistics that are neither Poisson nor Wigner. These blocks are identified by verifying that each is equivalent to a \gls{kt}. We establish a formal correspondence showing that the dynamical character of each block is determined by its effective \gls{kt} parameter, expressed in terms of the block dimension \(j\) and the parameter \(\tau_\subata\) of the
\gls{ata}
spin system. These findings are supported by numerical studies in which the
\gls{nns} distribution,
\gls{nns} spacing ratio,
and spectral rigidity are computed across blocks of varying sizes, confirming the predicted dynamics. These results highlight that the system exhibits not only integrable and chaotic dynamics but a continuous range of dynamical regimes across its symmetry blocks. Our approach based on symmetry subspaces should apply to generalised spin systems introduced in Ref~\cite{munozarias2021}. Different systems (with distinct parameters) exhibit different dynamics. We leave that for future work.

Our results imply that a single \gls{ata} device can host a continuum of
dynamical regimes without retuning its Hamiltonian: the regime is selected by
the symmetry sector (i.e., by state preparation within a given block). This
offers a route to engineer integrable, mixed, or chaotic behaviour via symmetry
selection rather than parameter tuning. The perturbation analysis further shows
that this diversity relies on symmetry resolution: once perturbations mix
sectors or become sufficiently strong, the block-dependent statistics collapse
to a single universal behaviour. Thus, symmetry-resolved spectra are both a
diagnostic of chaos and a control knob in experiments.

We emphasise that the present work is theoretical and does not provide a
direct comparison with experimental data. Nevertheless,
collective-spin/\gls{kt} dynamics and long-range Ising interactions relevant
here have been implemented experimentally in cold-atom and trapped-ion
platforms, so a direct benchmark is
feasible~\cite{chaudhury2009,Britton2012,Lu2025}. Moreover, our noise analysis
in Sec.~\ref{sec:perturbation} indicates that the key statistical signatures
persist for moderate perturbations, suggesting robustness against realistic
imperfections and motivating future experimental tests.

Finally, we verify that the variation in statistics from integrable to chaotic across different symmetry blocks persists until a perturbation with a norm of about 1 appears during evolution. With the increasing availability of multi-qubit quantum computers, these results can be experimentally tested and serve as a valuable platform for studying quantum chaos~\cite{Shepelyansky2001, Giraud2005, Seki2025}.

\ifdefined\EMBEDDEDINRESPONSE \else
\section*{Conflict of Interest} 

The authors declare that there are no conflicts of interest regarding the publication of this article.

\section*{Data Availability Statement} 

The code used to generate the results in this study is available on Zenodo at
\href{https://zenodo.org/records/18873357}{record 18873357}. The associated DOI
is \href{https://doi.org/10.5281/zenodo.18865339}{10.5281/zenodo.18865339}.

\section*{Acknowledgements} 
CP acknowledges support by UNAM-PAPIIT IG101324, SECIHTI CBF-2025-I-1548 and UNAM PASPA–DGAPA. DAA acknowledges support by NSERC of Canada, the Government of Alberta, projects DeQHOST APVV-22-0570, and QUAS VEGA 2/0164/25. C.P.\ acknowledges financial support from the Austrian Federal Ministry of Education, Science and Research via the Austrian Research Promotion Agency (FFG) through the project FO999921415 (Vanessa-QC) funded by the European Union{\textemdash}NextGenerationEU.

\bibliography{bibliography}

@book{Ozorio_1989, 
	place={Cambridge}, 
	series={Cambridge Monographs on Mathematical Physics}, 
	title={{H}amiltonian Systems: Chaos and Quantization}, 
	publisher={Cambridge University Press}, 
	author={Ozorio de Almeida, Alfredo M.}, 
	year={1989}, 
	collection={Cambridge Monographs on Mathematical Physics}
}

@article{karel,
  title = {Nonperiodic echoes from mushroom billiard hats},
  author = {Dietz, B. and Friedrich, T. and Miski-Oglu, M. and Richter, A. and Seligman, T. H. and Zapfe, K.},
  journal = {Phys. Rev. E},
  volume = {74},
  issue = {5},
  pages = {056207},
  numpages = {8},
  year = {2006},
  month = {Nov},
  publisher = {American Physical Society},
  doi = {10.1103/PhysRevE.74.056207},
  url = {https://link.aps.org/doi/10.1103/PhysRevE.74.056207}
}

@book{mehta2004random,
  title={Random Matrices},
  author={Mehta, Madan Lal},
  year={2004},
  publisher={Elsevier}
}

@article{Mehta1963,
  title = {Statistical theory of the energy levels of complex systems. {V}},
  volume = {4},
  ISSN = {1089-7658},
  url = {http://dx.doi.org/10.1063/1.1704009},
  DOI = {10.1063/1.1704009},
  number = {5},
  journal = {J. Math. Phys.},
  publisher = {AIP Publishing},
  author = {Mehta,  Madan Lal and Dyson,  Freeman J.},
  year = {1963},
  month = may,
  pages = {713–719}
}

@article{Atas2013,
  title = {Distribution of the ratio of consecutive level spacings in random matrix ensembles},
  author = {Atas, Y. Y. and Bogomolny, E. and Giraud, O. and Roux, G.},
  journal = {Phys. Rev. Lett.},
  volume = {110},
  issue = {8},
  pages = {084101},
  numpages = {5},
  year = {2013},
  month = {Feb},
  publisher = {American Physical Society},
  doi = {10.1103/PhysRevLett.110.084101},
  url = {https://link.aps.org/doi/10.1103/PhysRevLett.110.084101}
}

@article{Oganesyan2007,
  title = {Localization of interacting fermions at high temperature},
  author = {Oganesyan, Vadim and Huse, David A.},
  journal = {Phys. Rev. B},
  volume = {75},
  issue = {15},
  pages = {155111},
  numpages = {5},
  year = {2007},
  month = {Apr},
  publisher = {American Physical Society},
  doi = {10.1103/PhysRevB.75.155111},
  url = {https://link.aps.org/doi/10.1103/PhysRevB.75.155111}
}

@ARTICLE{Haake2010,
AUTHOR = {Haake, F.  and Kuś, M. },
TITLE   = {{K}icked top},
YEAR    = {2010},
JOURNAL = {Scholarpedia},
VOLUME  = {5},
NUMBER  = {11},
PAGES   = {10242},
DOI     = {10.4249/scholarpedia.10242},
NOTE    = {revision \#137061}
}

@book{Stckmann1999,
title = {Quantum Chaos: An Introduction},
ISBN = {9780511524622},
url = {http://dx.doi.org/10.1017/CBO9780511524622},
DOI = {10.1017/cbo9780511524622},
publisher = {Cambridge University Press},
author = {St\"{o}ckmann,  Hans-J\"{u}rgen},
year = {1999},
month = oct 
}

@article{Haake1987,
  author    = {Fritz Haake and Marek Ku{\'s} and Rainer Scharf},
  title     = {Classical and quantum chaos for a kicked top},
  journal   = {Z. Phys. B: Condens. Matter},
  year      = {1987},
  volume    = {65},
  number    = {3},
  pages     = {381--395},
  doi       = {10.1007/BF01303727}
}

@book{HaakeBook,
  author    = {Fritz Haake},
  title     = {Quantum Signatures of Chaos},
  publisher = {Springer},
  address   = {Berlin, Heidelberg},
  year      = {1991},
  edition   = {1st},
  doi       = {10.1007/978-3-662-02693-8}
}

@book{lichtenberg1992,
  author    = {Lichtenberg, A. J. and Lieberman, M. A.},
  title     = {Regular and Chaotic Dynamics},
  edition   = {2nd},
  publisher = {Springer-Verlag},
  address   = {New York},
  year      = {1992}
}

@book{ott2002,
  author    = {Ott, Edward},
  title     = {Chaos in Dynamical Systems},
  edition   = {2nd},
  publisher = {Cambridge University Press},
  address   = {Cambridge},
  year      = {2002}
}

@article{srednicki1994,
  author  = {Srednicki, Mark},
  title   = {Chaos and quantum thermalization},
  journal = {Phys. Rev. E},
  volume  = {50},
  pages   = {888--901},
  year    = {1994}
}

@article{rigol2008,
  author  = {Rigol, Marcos and Dunjko, Vanja and Olshanii, Maxim},
  title   = {Thermalization and its mechanism for generic isolated quantum systems},
  journal = {Nature},
  volume  = {452},
  pages   = {854--858},
  year    = {2008}
}

@article{chaudhury2009,
  author  = {Chaudhury, Souma and Smith, Anupam and Anderson, B. E. and Ghose, Shohini and Jessen, Poul S.},
  title   = {Quantum signatures of chaos in a kicked top},
  journal = {Nature},
  volume  = {461},
  pages   = {768--771},
  year    = {2009}
}

@book{sakurai1994,
  author    = {Sakurai, J. J.},
  title     = {Modern Quantum Mechanics (Revised Edition)},
  publisher = {Addison-Wesley},
  address   = {Reading, MA},
  year      = {1994}
}

@article{nandy2025,
  author  = {Nandy, Pratik and Matsoukas-Roubeas, Apollonas S. and Mart\'inez-Azcona, Pablo and Dymarsky, Anatoly and del Campo, Adolfo},
  title   = {Quantum dynamics in {K}rylov space: methods and applications},
  journal = {Physics Reports},
  volume  = {1125},
  pages   = {1--82},
  year    = {2025}
}

@article{bunimovich2001,
  author  = {Bunimovich, Leonid A.},
  title   = {Mushrooms and other billiards with divided phase space},
  journal = {Chaos},
  volume  = {11},
  number  = {4},
  pages   = {802--808},
  year    = {2001},
  doi     = {10.1063/1.1418763}
}

@article{lipkin1965,
  author  = {Lipkin, Harry J. and Meshkov, Nathan and Glick, Abraham J.},
  title   = {Validity of many-body approximation methods for a solvable model: ({I})},
  journal = {Nuclear Physics},
  volume  = {62},
  pages   = {188--198},
  year    = {1965}
}

@article{berry1977,
  author  = {Berry, M. V. and Tabor, M. and Ziman, J. M.},
  title   = {Level clustering in the regular spectrum},
  journal = {Proc. R. Soc. Lond. A},
  volume  = {356},
  pages   = {375--394},
  year    = {1977}
}

@article{bohigas1984,
  author  = {Bohigas, O. and Giannoni, M.-J. and Schmit, C.},
  title   = {Characterization of chaotic quantum spectra and universality of level fluctuation laws},
  journal = {Phys. Rev. Lett.},
  volume  = {52},
  pages   = {1--4},
  year    = {1984}
}

@article{berryrobnik1984,
  author  = {Berry, M. V. and Robnik, M.},
  title   = {Semiclassical level spacings when regular and chaotic orbits coexist},
  journal = {J. Phys. A: Math. Gen.},
  volume  = {17},
  pages   = {2413--2421},
  year    = {1984}
}

@article{Pineda2014,
  title = {Two-dimensional kicked quantum {Ising} model: dynamical phase transitions},
  volume = {16},
  ISSN = {1367-2630},
  url = {http://dx.doi.org/10.1088/1367-2630/16/12/123044},
  DOI = {10.1088/1367-2630/16/12/123044},
  number = {12},
  journal = {New Journal of Physics},
  publisher = {IOP Publishing},
  author = {Pineda,  C and Prosen,  T and Villaseñor,  E},
  year = {2014},
  month = dec,
  pages = {123044}
}

@article{Yoshimura2024,
  title = {Robustness of quantum chaos and anomalous relaxation in open quantum circuits},
  volume = {15},
  ISSN = {2041-1723},
  url = {http://dx.doi.org/10.1038/s41467-024-54164-7},
  DOI = {10.1038/s41467-024-54164-7},
  number = {1},
  journal = {Nat. Commun.},
  publisher = {Springer Science and Business Media LLC},
  author = {Yoshimura,  Takato and Sá,  Lucas},
  year = {2024},
  month = nov 
}

@article{Li2025,
  title = {Noise effects on the diagnostics of quantum chaos},
  volume = {111},
  ISSN = {2470-0029},
  url = {http://dx.doi.org/10.1103/PhysRevD.111.086008},
  DOI = {10.1103/physrevd.111.086008},
  number = {8},
  journal = {Phys. Rev. D},
  publisher = {American Physical Society (APS)},
  author = {Li,  Tingfei},
  year = {2025},
  month = apr 
}

@article{Nonnenmacher2003,
  title = {Spectral properties of noisy classical and quantum propagators},
  volume = {16},
  ISSN = {1361-6544},
  url = {http://dx.doi.org/10.1088/0951-7715/16/5/309},
  DOI = {10.1088/0951-7715/16/5/309},
  number = {5},
  journal = {Nonlinearity},
  publisher = {IOP Publishing},
  author = {Stéphane Nonnenmacher},
  year = {2003},
  month = jul,
  pages = {1685–1713}
}

@article{Ferrari2025,
  title = {Dissipative quantum chaos unveiled by stochastic quantum trajectories},
  author = {Ferrari, Filippo and Gravina, Luca and Eeltink, Debbie and Scarlino, Pasquale and Savona, Vincenzo and Minganti, Fabrizio},
  journal = {Phys. Rev. Res.},
  volume = {7},
  issue = {1},
  pages = {013276},
  numpages = {31},
  year = {2025},
  month = {Mar},
  publisher = {American Physical Society},
  doi = {10.1103/PhysRevResearch.7.013276},
  url = {https://link.aps.org/doi/10.1103/PhysRevResearch.7.013276}
}

@inproceedings{Shepelyansky2001,
  title = {Quantum chaos and quantum computers},
  url = {http://dx.doi.org/10.1142/9789812811004_0016},
  DOI = {10.1142/9789812811004_0016},
  booktitle = {Quantum Chaos Y2K},
  publisher = {WORLD SCIENTIFIC},
  author = {Shepelyansky,  D. L.},
  year = {2001},
  month = oct,
  pages = {112–120}
}

@article{Giraud2005,
  title = {Intermediate quantum maps for quantum computation},
  author = {Giraud, O. and Georgeot, B.},
  journal = {Phys. Rev. A},
  volume = {72},
  issue = {4},
  pages = {042312},
  numpages = {4},
  year = {2005},
  month = {Oct},
  publisher = {American Physical Society},
  doi = {10.1103/PhysRevA.72.042312},
  url = {https://link.aps.org/doi/10.1103/PhysRevA.72.042312}
}

@article{Seki2025,
  title = {Simulating {Floquet} scrambling circuits on trapped-ion quantum computers},
  author = {Seki, Kazuhiro and Kikuchi, Yuta and Hayata, Tomoya and Yunoki, Seiji},
  journal = {Phys. Rev. Res.},
  volume = {7},
  issue = {2},
  pages = {023032},
  numpages = {25},
  year = {2025},
  month = {Apr},
  publisher = {American Physical Society},
  doi = {10.1103/PhysRevResearch.7.023032},
  url = {https://link.aps.org/doi/10.1103/PhysRevResearch.7.023032}
}

@article{Kolmogorov1954,
  title={On conservation of conditionally periodic motions for a small change in {H}amilton's function},
  author={A. N. Kolmogorov},
  journal={Dokl. Akad. Nauk SSSR},
  year={1954},
  volume={98},
  pages={527-530}
}

@article{Arnold1963,
  title = {Proof of a theorem of {A}.~{N}. {Kolmogorov} on the invariance of quasi-periodic motions under small perturbations of the {Hamiltonian}},
  volume = {18},
  ISSN = {1468-4829},
  url = {http://dx.doi.org/10.1070/RM1963v018n05ABEH004130},
  DOI = {10.1070/rm1963v018n05abeh004130},
  number = {5},
  journal = {Russ. Math. Surv.},
  publisher = {Steklov Mathematical Institute},
  author = {Arnol’d,  V I},
  year = {1963},
  month = oct,
  pages = {9–36}
}

@article{Passarelli2025,
  title = {Chaos and magic in the dissipative quantum kicked top},
  volume = {9},
  ISSN = {2521-327X},
  url = {http://dx.doi.org/10.22331/q-2025-03-05-1653},
  DOI = {10.22331/q-2025-03-05-1653},
  journal = {Quantum},
  publisher = {Verein zur Forderung des Open Access Publizierens in den Quantenwissenschaften},
  author = {Passarelli,  Gianluca and Lucignano,  Procolo and Rossini,  Davide and Russomanno,  Angelo},
  year = {2025},
  month = mar,
  pages = {1653}
}

@article{munozarias2021,
  title = {Nonlinear dynamics and quantum chaos of a family of kicked $p$-spin models},
  author = {Mu\~noz-Arias, Manuel H. and Poggi, Pablo M. and Deutsch, Ivan H.},
  journal = {Phys. Rev. E},
  volume = {103},
  issue = {5},
  pages = {052212},
  numpages = {21},
  year = {2021},
  month = {May},
  publisher = {American Physical Society},
  doi = {10.1103/PhysRevE.103.052212},
  url = {https://link.aps.org/doi/10.1103/PhysRevE.103.052212}
}

@article{carollo2021,
  title = {Exactness of Mean-Field Equations for Open {D}icke Models with an Application to Pattern Retrieval Dynamics},
  author = {Carollo, Federico and Lesanovsky, Igor},
  journal = {Phys. Rev. Lett.},
  volume = {126},
  issue = {23},
  pages = {230601},
  numpages = {6},
  year = {2021},
  month = {Jun},
  publisher = {American Physical Society},
  doi = {10.1103/PhysRevLett.126.230601},
  url = {https://link.aps.org/doi/10.1103/PhysRevLett.126.230601}
}

@article{Britton2012,
  title = {Engineered two-dimensional {Ising} interactions in a trapped-ion quantum simulator with hundreds of spins},
  volume = {484},
  ISSN = {1476-4687},
  url = {http://dx.doi.org/10.1038/nature10981},
  DOI = {10.1038/nature10981},
  number = {7395},
  journal = {Nature},
  publisher = {Springer Science and Business Media LLC},
  author = {Britton,  Joseph W. and Sawyer,  Brian C. and Keith,  Adam C. and Wang,  C.-C. Joseph and Freericks,  James K. and Uys,  Hermann and Biercuk,  Michael J. and Bollinger,  John J.},
  year = {2012},
  month = apr,
  pages = {489–492}
}

@article{Lu2025,
  title = {Implementing arbitrary {Ising} models with a trapped-ion quantum processor},
  author = {Lu, Yao and Chen, Wentao and Zhang, Shuaining and Zhang, Kuan and Zhang, Jialiang and Zhang, Jing-Ning and Kim, Kihwan},
  journal = {Phys. Rev. Lett.},
  volume = {134},
  issue = {5},
  pages = {050602},
  numpages = {7},
  year = {2025},
  month = {Feb},
  publisher = {American Physical Society},
  doi = {10.1103/PhysRevLett.134.050602},
  url = {https://link.aps.org/doi/10.1103/PhysRevLett.134.050602}
}

@article{Bapst2012,
  title = {On quantum mean-field models and their quantum annealing},
  volume = {2012},
  ISSN = {1742-5468},
  url = {http://dx.doi.org/10.1088/1742-5468/2012/06/P06007},
  DOI = {10.1088/1742-5468/2012/06/p06007},
  number = {06},
  journal = {J. Stat. Mech: Theory Exp.	},
  publisher = {IOP Publishing},
  author = {Bapst,  Victor and Semerjian,  Guilhem},
  year = {2012},
  month = jun,
  pages = {P06007}
}
\fi

\end{document}